# Ultra-high frame rate digital light projector using chipscale LED-on-CMOS technology

NAVID BANI HASSAN,[1] FAHIMEH DEHKHODA,[2] ENYUAN XIE,[1] JOHANNES HERRNSDORF,[1*] MICHAEL J. STRAIN,[1] ROBERT HENDERSON[2] AND MARTIN D.DAWSON[1]

[1]*Institute of Photonics, Department of Physics, University of Strathclyde, Technology and Innovation Centre, Glasgow G1 1RD, UK*
[2]*School of Engineering, Institute for Integrated Micro and Nano Systems, University of Edinburgh, Edinburgh, EH9 3JL, UK*
*\*johannes.herrnsdorf@strath.ac.uk*

**Abstract:** Digital light projector systems are crucial components in applications including computational imaging, fluorescence microscopy and highly parallel data communications. Here we present a chip-scale projector system based on emissive micro-LED pixels directly bonded to a smart pixel CMOS drive chip. Enabled by the high modulation bandwidth of the LED devices, the 128×128 pixel array can project binary patterns at up to 0.5 Mfps and toggle between two stored frames at MHz rates. The projector has a 5-bit grayscale resolution that can be updated at rates up to 83 kfps, and can be held in memory as a constant bias for the binary pattern projection. Finally, the projector can be operated in a pulsed mode, with individual pixels emitting pulses down to a few nanoseconds in duration. Again, this mode can be used in conjunction with the high-speed spatial pattern projection. The design of the smart pixels and LED devices are presented along with measurements of each mode of operation. As a demonstration of the data throughput achievable with this system we present an optical camera communications application, exhibiting data rates of >5 Gbps, over three orders of magnitude improvement on current demonstrations.

## 1. Introduction

Digital light projection (DLP) systems are now a well-established technology, with advances driven by the demands of cinematic and consumer markets. Systems with kilo-frames per second (fps) update rates and pixel counts in the mega-pixel range are widely available serving both consumer video and scientific applications. Commercial developments are strongly driven by demand for high pixel count devices for digital projection and displays, though a number of other application areas have emerged including optical wireless communications [1], data through display [2-5], visible light positioning [6], single-pixel and 3D imaging [7-12], fluorescence lifetime imaging [13], and optogenetic stimulation [14, 15]. These applications benefit from, or even require, pixel modulation rates well beyond the typical frame rates of visual displays.

Current commercially available DLP systems are mainly based on two forms of underpinning technology, Liquid Crystal Displays (LCDs) and Digital Micro-Mirror Displays (DMDs) [16, 17]. LCDs use electronic control of nematic liquids, commonly sandwiched between polarization films, to control the absorption or phase change of light through a pixel [17]. DMDs are based on Micro Electro-Mechanical Systems (MEMS) devices, where the reflective surface of a pixel can be orientated to pass or block incident light through the projection path [16]. DMDs are particularly attractive due to their pixel switching times and compatibility with drive electronics. Whilst commercial development has enabled the impressive resolution of these systems, their frame rate is limited by the physical motion required to switch pixels and the data-handling capacity required to update Mpixel frames. Furthermore, grayscale operation of DMD displays uses duty-cycle control of pixel switching [16], consequently reducing the maximum achievable frame-rate of the device inversely with the number of gray levels required.

In parallel to these developments, advances in semiconductor materials, and in particular the III-Nitrides, has enabled the realization of high brightness LED pixel emitters with dimensions in the mm to µm range [18]. The spectral quality, brightness and potential for integration with electronic backplanes, has seen deployment of these devices in displays ranging in dimension from mobile phone interfaces to super-large screen installations for advertising or theatrical installations. In addition to their emissive properties, LED pixels have another key advantage to offer in DLP systems, namely their high modulation bandwidth, with individual devices demonstrated up to GHz rates [19], and the potential for short pulse emission in the sub-nanosecond regime [20].

The planar format of LED pixel arrays allows them to be directly interfaced with electronic drive chips, for example through flip-chip bonding [21]. Arrays of LEDs with individual pixel control electronics represent a new form of compact micro-display/projector [18]. In previous work we have reported on LED arrays with 16×16 and 10×40 pixel arrays, integrated onto complementary metal-oxide semiconductor (CMOS) driver chips [6, 22, 23], along with proof-of-concept demonstrations including Gbps data communications [6, 24, 25], time of flight ranging [8], spatial multiplexing and navigation [6]. These initial integrated chip systems were designed to allow probing of multiple modes of operation and were therefore very flexible devices but limited in total pixel number and effective frame rate, the latter being 30 kfps in the best case.

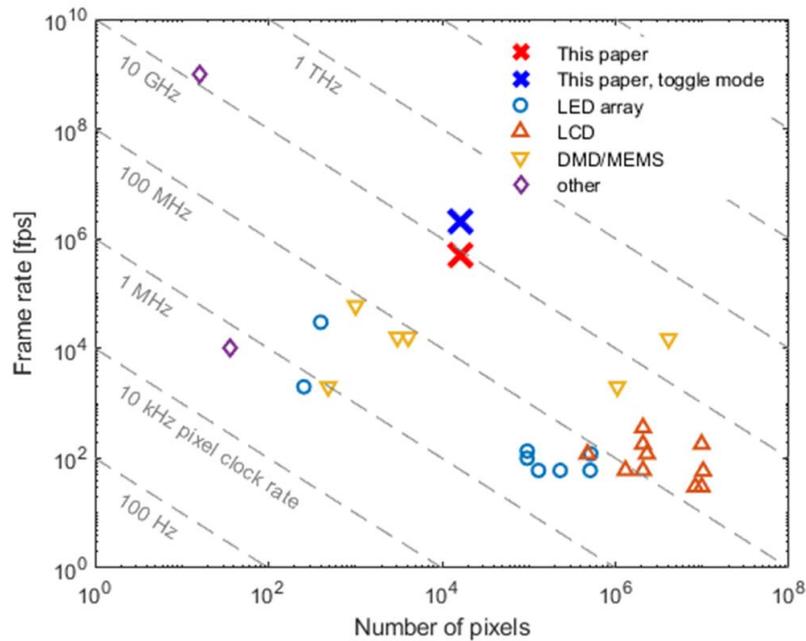

Fig. 1. Comparison of the LED array presented here with other 2D programmable pattern sources [3-6, 22, 26-28], with pixel count and frame rate as parameters.

In this work we present a chip-scale LED-on-CMOS DLP using a 128×128 pixel array optimized for high speed spatio-temporal pattern projection. The device is designed with an electronic interface to allow pattern projection at up to 0.5 Mfps, 5 nano-second pulsed operation, grayscale pixel intensity and pattern toggling in excess of 2 MHz using on-chip memory. Furthermore, these individual modes can be operated simultaneously, for example allowing image projection at video rates, together with high-speed data transmission.

Fig. 1 compares the system presented here to other DLP technologies including the early stage research prototypes from references [3-6, 22, 26-28] and the commercially available devices listed in the supplementary information (S2). The dotted lines on the figure represent effective pixel clock rate, calculated as the frame rate multiplied by the number of available pixels. It can be seen that compared to established technologies based on LCD, DMD/MEMS and previous LED arrays, the LED array presented here provides an order of magnitude improvement in frame rate at a competitive pixel count. The highest frame rate reported for a 2D light modulator to our knowledge is 1 Gfps, achieved with a 4×4 array at a wavelength of 1550 nm and providing phase modulation only [28]. Even though this phase modulator has been included in Fig. 1 (diamond data point upper left), its low pixel count and operating principles make it incompatible with the application space of the high-density intensity modulation in the visible provided by the micro-LED array. There are also 1D pixel array approaches that achieve update rates near the Mfps mark, albeit at a significantly lower pixel count [27]. There is to our knowledge no other technology that currently provides Mfps rates at >10k pixels, or that combines patterned projection at these rates with nanosecond pulsing capability and grayscale intensity control.

This paper presents the performance of the chip in each operation mode, in addition to an example application in high speed optical camera communications (OCC). This device has been developed as a technology platform and we believe it will provide a significant advantage for applications currently using DLP technology, including fluorescence lifetime imaging and machine vision systems.

## 2. DLP hardware

The main design features of the system are presented in Table 1 and each aspect will be discussed in further detail below.

**Table 1. Design Specifications**

| Parameter | Value | Additional Detail |
|---|---|---|
| Array dimensions | 128×128 | |
| Pixel pitch | 50 µm | |
| Pixel active area | 30×30 µm$^2$ | |
| Pixel fill factor | 36% | Light emitting area |
| Frame rate | 0.5 Mfps | Limited by 8 Gb/s digital electronic interface between external FPGA controller and LED chip |
| Frame update mode | Global / Rolling | Both available irrespective of frame rate |
| Pattern toggling rate | 2 MHz | 8 MHz possible at compromised duty cycle/intensity |
| Frame rate (gray scale) | 83 kfps | Limited by 8 Gb/s digital electronic interface between external FPGA controller and LED chip |
| Number of gray values | 32 | 5 bit |
| LED current | 87 µA | Single pixel, highest gray level setting |
| Optical output power | 22 µW | Single pixel, highest gray level setting |
| Display brightness | 2.8×10$^4$ cd/m$^2$ | LEDs emitting at 450 nm wavelength |
| Maximum pulse repetition rate | 100 MHz | |
| Minimum pulse duration | 4 ns | Limited by external controller |
| Pulse energy | 0.08 pJ | Single pixel at 4 ns, 20 µW peak power |
| | 17 pJ | Pattern with 1000 active pixels at 4 ns, 1.5 A peak current |

## 2.1. Device fabrication

The LED array was fabricated using a c-plane GaN-on-sapphire wafer with standard LED fabrication processes which follow the design guidelines in reference [23]. The LED array was flip-chip bonded onto the CMOS driver chip using an indium-based bonding process, and the flip-chip bonded device was packaged and wire-bonded into a ceramic package, which can be inserted into a socket on a custom-made printed circuit board. The digital control signals to the CMOS chip are supplied by an Opal Kelly XEM7310-A200 field-programmable gate array (FPGA). Fig. 2 shows microscope images of the fabricated CMOS driver chip and LED array system.

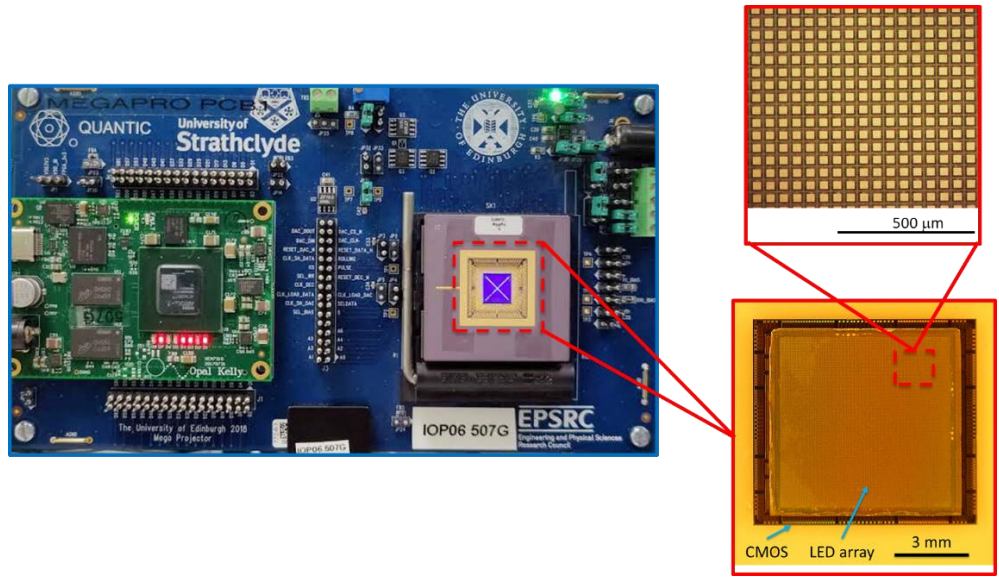

Fig. 2. Photograph of the mounted LED projector on CMOS chip, co-packaged with an FPGA controller. The projector is displaying a diagonal cross pattern. Inset images show the mounted LED array on CMOS chip and a zoomed view of the LED pixel contact pads, imaged through the sapphire substrate.

A total of 6 LED chips were fabricated for this work, all of which emit at 450 nm. In principle, any wavelength supported by the GaN material system can be used, i.e. 280 nm – 520 nm. The critical step in the system fabrication flow is the flip-chip bonding of the LED chip to the CMOS driver chip. The functional optical pixel yield for the 6 completed systems ranged between 30% and 86%, with 4 of the 6 devices having a yield above 77%, representing >12.5x$10^3$ active pixels. The pixel yield values are representative of the research lab fabrication processing and can be expected to be much improved using industrial foundry processing and quality control. The results presented in the manuscript correspond to the device with 86% pixel yield, i.e. >14k active pixels.

## 2.2. Smart pixel driver details and functionality

The multiple modes of operation of the projector chip, as detailed in Table 1, are enabled by the advanced functionality of each pixel driver, which comprises a 5-bit current digital to analogue converter (DAC), a separate current source with 2 mA driving capability, a 2-bit in-pixel memory for high-speed pattern toggling, and control signals to enable different operation modes. Fig. 3 shows a schematic circuit diagram of these elements and their connections. The chip infrastructure that supplies signals and power to each pixel is detailed in the supplementary information (S1). A simplified timing diagram is presented in Fig. 4.

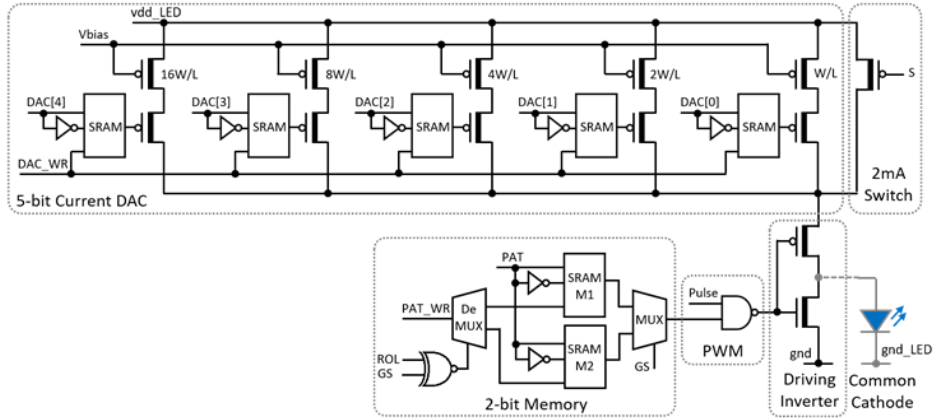

Fig. 3. Simplified schematic of the driver circuit for one pixel, showing 5-bit DAC, in-pixel memory and control signals.

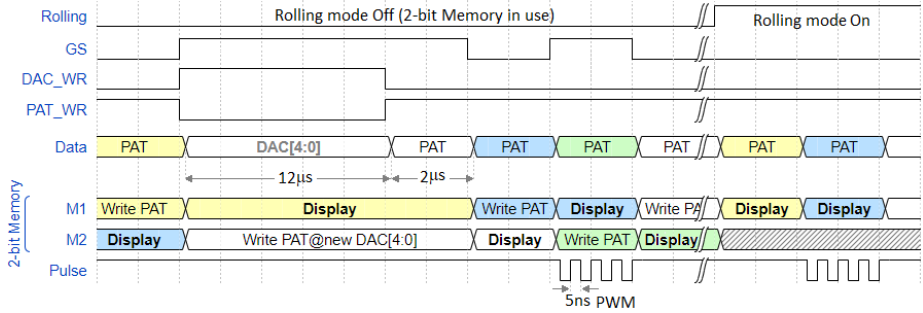

Fig. 4 Simplified timing diagram of the digital control signals supplied to the CMOS by an FPGA.

The optical output power of a single pixel driven by the DAC is shown in Fig. 5(a). The DAC was designed to provide currents ranging from 0 µA (gray level 0) to 87 µA (gray level 31) and a good linear behavior is observed in the output power apart from a small kink between gray levels 15 and 16 which is a well-known effect in this type of CMOS current DAC [29] The current levels per pixel were chosen in order to reduce electrical cross-talk [23] and keep the maximum current across the entire chip on the order of 1 A, which is a value that was found to be sustainable by previous generations of CMOS-controlled micro-LED arrays [23]. An example of a gray-scale image (the QuantIC project logo) from the device is shown in Fig. 5(b). This image was taken using a monochrome camera to better illustrate the gray levels.

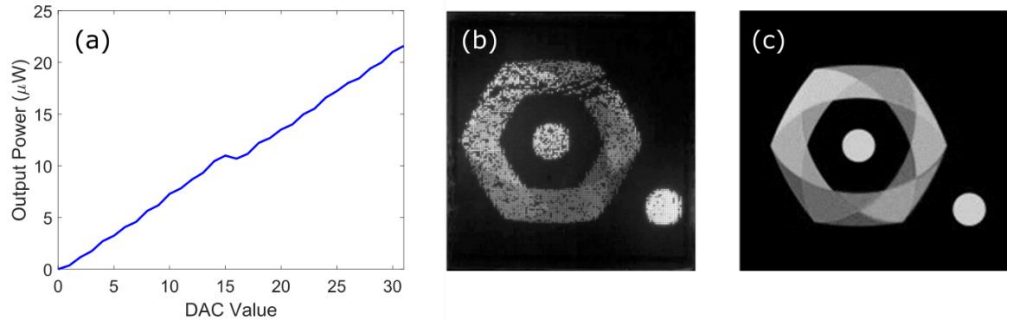

Fig. 5. (a) Optical output power of a single pixel for different DAC settings, (b) grayscale image from the LED projector chip using the DAC gray levels (c) digital version of the image projected in (b).

## 3. Demonstration of system operation modes

As indicated above in Table 1, there are 4 different modes of pattern switching accessible using the custom driver chip:

1. Rolling pixel update
2. Global pixel update
3. Global pixel toggling
4. Nano-second pulse mode

Modes 1 and 2 can be accessed for both binary and grayscale pattern projection. Mode 3 is accessible as a binary switching pattern. In all cases, grayscale information can be pre-stored on-chip and the binary pattern projection, toggling or pulsing, can be applied on top of this pre-existing bias setpoint. Details of each operation mode are presented in the following sections.

### 3.1 High speed pattern projection and optical camera communications

For binary pixel on-off state information, there are two update modes, rolling and global pixel update. For these cases, a single, full-frame, binary pattern can be loaded onto the chip within 2 µs, demonstrating the capability of pattern loading at 0.5 Mfps at full resolution. In the rolling pixel mode, each pixel is updated individually in sequence, with the value held until the next update. In the global pixel mode the full frame is refreshed in one shot once all of the pattern data is loaded.

In addition to the binary pixel on-off data, grayscale values can be sent to each pixel in the array. Significantly, as compared with DMD devices, the grayscale value corresponds to the pixel drive current and is not a duty cycle effect that relies on multiple binary frames for implementation. The gray scale pattern that is held by the 5-bit in-pixel DAC, can be loaded within 12 µs, corresponding to 83 kfps. As detailed in the supplementary information (S1), input data can be selectively written to the binary pixel state or the DAC, corresponding to the switch control signal. Therefore, a grayscale image can be stored on the chip and used as a set-point for rapid pixel on-off modulation.

To demonstrate the potential of this high speed binary projection, we used it in an optical camera communications (OCC) application. To establish an OCC link, the LED array was directly imaged onto a Photron UX100 high speed camera using a Tamron SP AF Macro 90mm F/2.8 Di lens. A collection of $10^4$ pseudo-random patterns were stored on the FPGA memory and were loaded in sequence onto the LED projector chip as described above. The system was operated in global update mode to enable synchronization of the pattern data with the camera frame rate. Full details of the experimental setup and data handling are presented in the supplementary information (S3). The main bottleneck in characterizing this mode of operation is in the camera frame rate. The Photron camera has a maximum full frame rate of 4 kfps for an imaging array of 1280 x 1024 pixels. Therefore, to match the speed of the LED projector the camera had to be operated in a reduced pixel area mode. With an active area of 640 x 8 pixels, the camera can operate at 0.8 Mfps. We randomly selected areas of the projector to image onto the camera slice and operated the projector at 0.25 Mfps and 0.4 Mfps, the latter representing the highest projector rate that the camera could temporally resolve at the Nyquist sampling limit. Fig. 6 shows eye diagrams and detected symbol level histograms for the $10^4$ frame data sets captured for both operating rates.

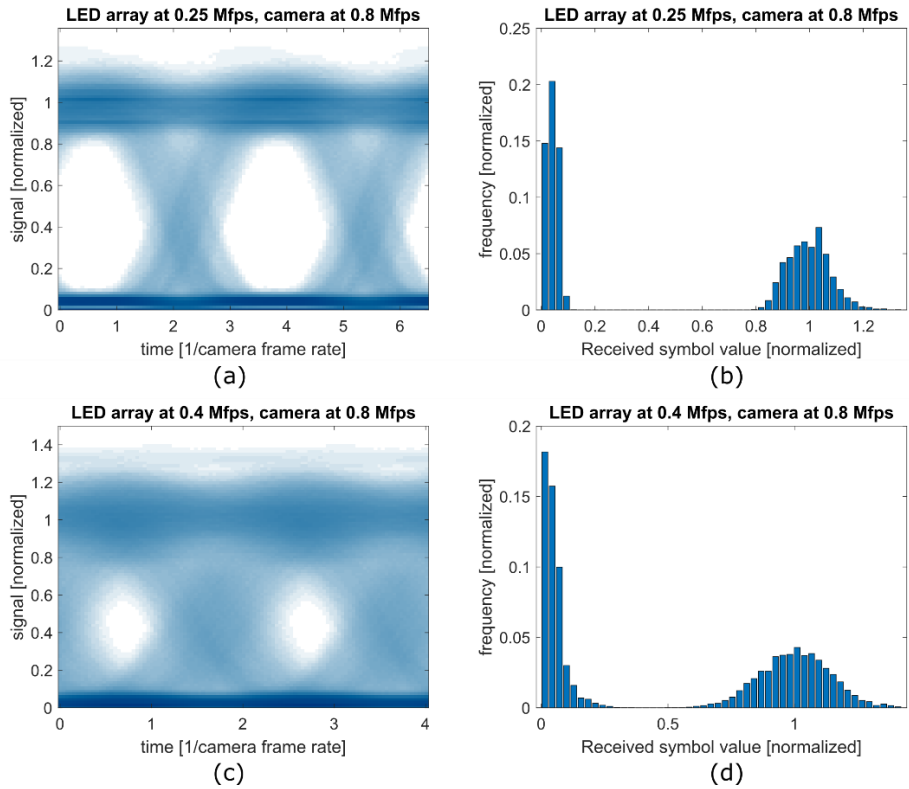

Fig. 6. a) Measured eye diagram for $10^4$ transmitted frames at 0.25 Mfps rate and (b) associated signal level histogram. (c) Measured eye diagram at 0.4 Mfps and (d) associated signal level histogram.

The eye diagram for the 0.25 Mfps case is very clear and represents error free transmission (for the $10^4$ sample limit). The signal level histogram shows clear separation of the 0 and 1 bit levels. For the 0.4 Mfps projector rate the eye diagram is more closed than for the 0.25 Mfps case. This is due to the sampling limit of the camera rather than intensity level variations in the projector, with clear time jitter measurement limitations present in the eye diagram. The symbol level histogram shows very clear distinction between the 0 and 1 bit levels and again, error free data transmission up to $10^4$ transmitted frames was achieved. Taking into account the 86% pixel yield of the chip, aggregate data rates of 3.5 Gbps and 5.6 Gbps were transmitted for the 0.25 Mfps and 0.4 Mfps frame rates, respectively.

## 3.2 High-Speed pattern toggling

Using the 2-bit in-pixel memory function, it is possible to exceed the switching rate of the data projection modes detailed above, in a two pattern toggling mode. In this mode, two independent frames can be stored using the in-pixel memory and switched using a global clock signal. As before, the switching performance of the frame toggling mode has been assessed by imaging the patterns onto a high-speed camera (Photron UX100). To achieve the maximum frame rate of the camera, the 640×8 pixel sub-region of the available frame area was used. We selected two pattern sets for the projector system toggling tests: (1) checkerboard switching to all pixels off (C2B) and (2) two spatially complementary checkerboard patterns (C2C). Fig. 7 shows raw captured images, where each image shows a 20×2 pixel sub-section of the LED array which covers the field of view of the camera. Fig. 7(a) and (b) are temporally adjacent image frames at 0.8 Mfps for the C2B patterns. Similarly, Fig. 7(c) and (d) are temporally adjacent frames of the C2C patterns. The LED array was toggling at 0.8 Mfps and these frames are representative frames from a video where the LED array and camera capture were in-phase with one another. The full switching limits of the array are beyond 0.8 Mfps and therefore cannot be resolved by the Photron UX100 camera.

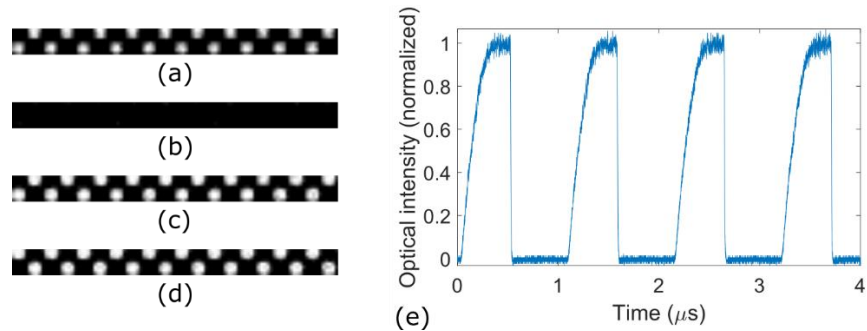

Fig. 7. Representative LED projection frames captured with the Photron camera at 0.8 Mfps. Visible is the partial footprint of the last two columns of the LED chip: (a) frame #1 C2B, (b) frame #2 C2B, (c) frame #1 C2C, and (d) frame #2 C2C. (e) Measured photo-signal from global array toggling at 2 Mfps.

To determine the pattern switching limits in toggling mode, the array was toggled between all pixels off and a pattern with all pixels on at the highest DAC setting of 31, and the optical output was measured with a photomultiplier tube (PMT). Fig. 7(b) shows the recorded waveform at a toggling rate of 2 Mfps, showing full on/off switching with a rise time of 315 ns and a fall time of 8 ns. Above 2 Mfps a reduction in duty cycle and peak power was observed and at 8 Mfps the duty cycle had reduced to 25%. This behavior was independent of the number of pixels that were switched.

## 3.3. Nanosecond pulsed mode

Compared with other display technologies the micro-LED platform has the unique capability of emitting nanosecond or even sub-nanosecond optical pulses [20]. In the device presented here, this capability is accessed by selecting a 2mA drive line in the pixel, which enables a higher LED drive current than the DAC and is controlled by a separate global control signal which can be switched on a nanosecond timescale. Pixels are enabled to respond to the global pulse signal using the 2-bit memory, thus allowing spatial patterns to be pulsed, where the patterns can be updated at the same rate as detailed previously.

The nanosecond pulsing function was assessed by loading patterns into the 2-bit memory with different numbers and distributions of active pixels, and pulsing them with electronic pulse widths between 5 ns and 100 ns at repetition rates between 10 MHz and 100 MHz. The optical pulses were resolved using a PMT connected to an oscilloscope with 1 GHz bandwidth, and the spatial patterns were verified using a slow response optical camera.

Fig. 8(a) shows a representative spatial projection pattern that is being driven in nanosecond pulsed mode. Representative temporal waveforms recorded by the PMT at 10 MHz repetition rate, 5 ns pulse duration, and with the number of active pixels as a parameter, are shown in Fig. 8(b), highlighting the consistent pulse shape over a wide range of active device numbers. The number of active pixels is limited to ~5000 in these measurements since optical emission was significant reduced beyond this, as detailed below. The electrical drive pulse width is 5 ns, while the emitted optical pulse width is ~4 ns due to the rise and fall times of the LED pixel. As shown in Fig. 8(c), the measured optical pulse width reduces as a function of increasing active pixel count. The effect is most pronounced for the 10 MHz repetition rate measurements, where the pulse width is dominated by the rise and fall times of the LED. The capacitive loading associated with increasing numbers of active pixels leads to a delay in the optical rise time and thus to pulse shortening. At higher pulse repetition rates, approaching a 50 % duty cycle, the FWHM pulse width is more stable, but is attended by a reduction in extinction ratio.

To assess the effect of the distribution of the global pulse signal in the nanosecond time regime, the timing delay of emitted pulses across the array was measured. Local arrays of the LED chip were imaged onto the PMT and the received signal was compared to the supply pulse trigger signal to measure relative delay. Fig. 8(e) and (f) show the relative time delay between optical pulses emitted by individual pixels at various locations on the chip, arranged by column or row respectively. The time delay between columns on chip does not show a clear trend, with a scatter < 0.1 ns. There is however a clear time delay trend with respect to the chip rows. The maximum delay scatter across the full chip is still < 0.7 ns, significantly lower than the pulse width.

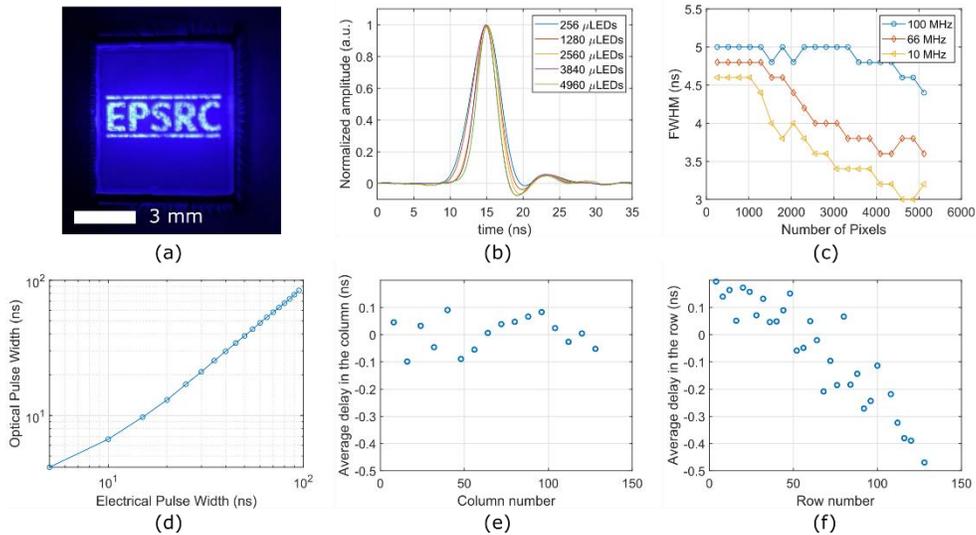

Fig. 8 a) optical micrograph of a pattern displayed in nanosecond pulsed mode, b) optical waveform of 5 ns long pulses as a function of number of pulsing pixels, c) optical pulse duration as a function of the number of pixels for a 5 ns input pulse, d) optical pulse duration as a function of electrical input pulse duration, e) Relative delay of the pulses emitted by individual pixels within one row f) Relative delay of the pulses emitted by individual pixels within one column.

The emitted optical pulse energy depends on the number of active pixels as shown in Fig. 9, where the pulse duration was 5 ns and the repetition rate was 10 MHz. From Fig. 9(a) the total aggregated pulse energy increases as a function of active pixel number, then goes through a turning point, reducing as further pixels are added, beyond ~1000 active pixels. At the maximum emission energy, the average chip current is 0.07 A, corresponding to a peak switching current of 1.5 A. As the number of active pixels increases, the average and peak currents increase to 0.12 A and 2.4 A respectively. The corresponding decrease in optical pulse energy at these levels suggests that most of the electrical energy is being dissipated in capacitive charging, as highlighted in Fig. 9(b), which shows the per pixel pulse energy as a function of active pixel number.

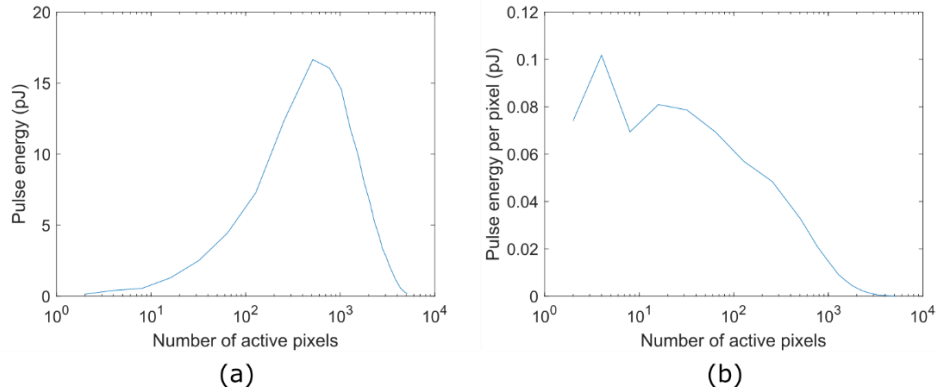

Fig. 9. (a) Total chip pulse energy for 5 ns long pulses and average current drawn by the device as a function of the number of active pixels at 10 MHz, and (b) pulse energy per pixel as a function of number of active pixels.

## 4. Discussion

The DLP system presented makes use of LED emitters for their high brightness, direct electronic emission control and potential for high bandwidth modulation. The particular pixels in this system have emitter dimensions of $30\times30$ µm$^2$, and devices with comparable footprints have been shown to be capable of modulation bandwidths in excess of 100 MHz when driven at high current densities in excess of 1 kA/cm$^2$ [25]. Although targeting high speed frame rates, there is a limitation associated with how these LED pixels are driven. Our devices typically operate at current densities of ~10 A/cm$^2$, chosen in order to limit the overall current drawn by the device and to mitigate electrical cross-talk and current-crowding effects that have been observed at higher current densities in the past [23]. The latter is a particularly important issue for high pixel count projectors where uniform optical emission is required. At this lower current density, the LEDs still have bandwidths of a few MHz, and are thus intrinsically faster than the established LCD and DMD technologies which provide 10's of kHz. Furthermore, the grayscale emission control that can be operated concurrently with the binary pattern projection and short pulse modes does not degrade the effective system frame rate as is the case in DMD devices. Nevertheless, in applications requiring higher pulse energies or switching speeds, there are clear design trade-offs between total pixel number, the spatial density of patterns required and the power handling of the chip that could be optimized in these cases.

We present a 128×128 pixel array device with a fabrication pixel yield of 86% in the best case. This is not a fundamental limit on the integration technology and is a result of the prototyping fabrication processes used. Using established foundry fabrication routes and opto-electronic integration there is no reason that future commercial devices could not reach production compatible yields. In fact, the components are compatible with current industrial process in micro-displays and electronic chip fabrication.

The nanosecond pulse measurements presented were limited to 5 ns minimum duration. These timescales are compatible with time-correlated single-photon counting (TCSPC) applied for example to fluorescence lifetime imaging or time-of-flight ranging [8, 13]. Again, if particular applications require shorter pulse durations or higher per pixel energies, these are achievable by optimizing both the CMOS driver and LED pixel designs, but were taken as a reasonable design target in this multi-functional demonstrator chip. The high spatial resolution nanosecond pulse operation in addition to high speed pattern generation has clear applications in fluorescence microscopy and computational imaging.

Finally, whilst the capabilities of the chip have been discussed in the main as a source for imaging and sensing applications, it should also be noted that they also present significant potential as high-speed pattern-programmable excitation sources for a wide range of optically pumped physical effects, for example in spatio-temporally resolved fluorescence lifetime imaging.

## 5. Conclusion

By employing micro-LED emitters as elements in digital light projection systems, a number of complementary modes of operation can be achieved, with frame update rates significantly higher than current mechanically based systems. We have presented a smart pixel control CMOS drive chip directly bonded to an emissive micro-LED pixel chip with an active pixel count >14k. By employing the direct current control of each pixel element, the array can be used to project patterns at rates up to 0.5 Mfps, and over 2 MHz in a two pattern toggling mode. This represents an order of magnitude improvement over binary pattern projection using DMD displays. Furthermore, in-pixel memory enables direct grayscale control of emission levels concurrently with binary pattern switching, without degrading frame rates. Finally, the chip can be operated with a rolling or global pixel update and in a nanosecond pulse mode with sub-nanosecond delay across the full array. We illustrated the potential of this system in an optical camera communications application, demonstrating data transmission rates greater than 5 Gbs, over three orders of magnitude over current state of the art in OCC. The operating characteristics of this system make it an attractive source for future applications in computational imaging, microscopy and spatio-temporally controlled optical pumping of physical systems.

**Funding.** Engineering and Physical Sciences Research Council (EP/M01326X/1, EP/T00097X/1, EP/S001751/1), Royal Academy of Engineering (Research Chairs and Senior Research Fellowships).

**Acknowledgments.** F. Dehkhoda and N. Bani Hassan contributed equally to this work. We thank Fraunhofer IZM for their help with flip-chip bonding. The authors thank Graeme Johnstone for help with the toggle mode characterization.

**Disclosures.** The authors declare no conflicts of interest.

**Data availability.** Data underlying the results presented in this paper are available in Ref. [30].

**Supplemental document.** See Supplement 1 for supporting content.

# Ultra-high frame rate digital light projector using chipscale LED-on-CMOS technology - Supplementary Information

NAVID BANI HASSAN,[1] FAHIMEH DEHKHODA,[2] ENYUAN XIE,[1] JOHANNES HERRNSDORF,[1*] MICHAEL J. STRAIN,[1] ROBERT HENDERSON[2] AND MARTIN D. DAWSON[1]

[1]*Institute of Photonics, Department of Physics, University of Strathclyde, Technology and Innovation Centre, Glasgow G1 1RD, UK*
[2]*School of Engineering, Institute for Integrated Micro and Nano Systems, University of Edinburgh, Edinburgh, EH9 3JL, UK*
*\*johannes.herrnsdorf@strath.ac.uk*

### 6. CMOS smart pixel design

The detailed block diagram of the CMOS chip is illustrated in Figure S1. This CMOS backplane includes 128×128 driver cells to drive 128×128 micro-LEDs which are bump-bonded to their dedicated driver cell. The chip was implemented in 0.35 $\mu m$ CMOS technology from Austria Microsystems and occupies 7.66 mm×8.33 mm. The pixel pitch is 50µm with 30µm x 30µm active area for the micro-LED mesa. Each pixel driver is connected to the anode contact of micro-LED. All micro-LEDs have a common cathode labeled as gnd_LED which is separated from the ground terminal (gnd) of the driver circuit. This allows achieving higher optical output power by pulling gnd_LED below the global ground level and biasing the micro-LEDs with higher voltage. The common cathode is connected to a wide cathode ring around the pixel grid and multiple pads. Two voltage domain are used on the chip, 3.3V (vdd) for the control logic and 5V (vdd_LED) inside the pixels to drive the micro-LEDs.

A power grid is adopted on the pixel grid to evenly distribute the power supply and ground lines in different rows and columns. The chip takes benefit of in-pixel memory unit to store the LED's ON/OFF state and therefore the entire LED array can be illuminated globally. Independent intensity control of each LED using a local 5-bit current DAC is another advantage of the driver chip which can produce grayscale patterns. This feature can also compensate the optical power mismatch between pixels arising from the process variations or bump bonding issues. A global PWM techniques is also included to pulse the micro-LEDs with a minimum pulse duration of 5ns when they are driven with 2mA current. This is achievable through the input signal labeled Pulse.

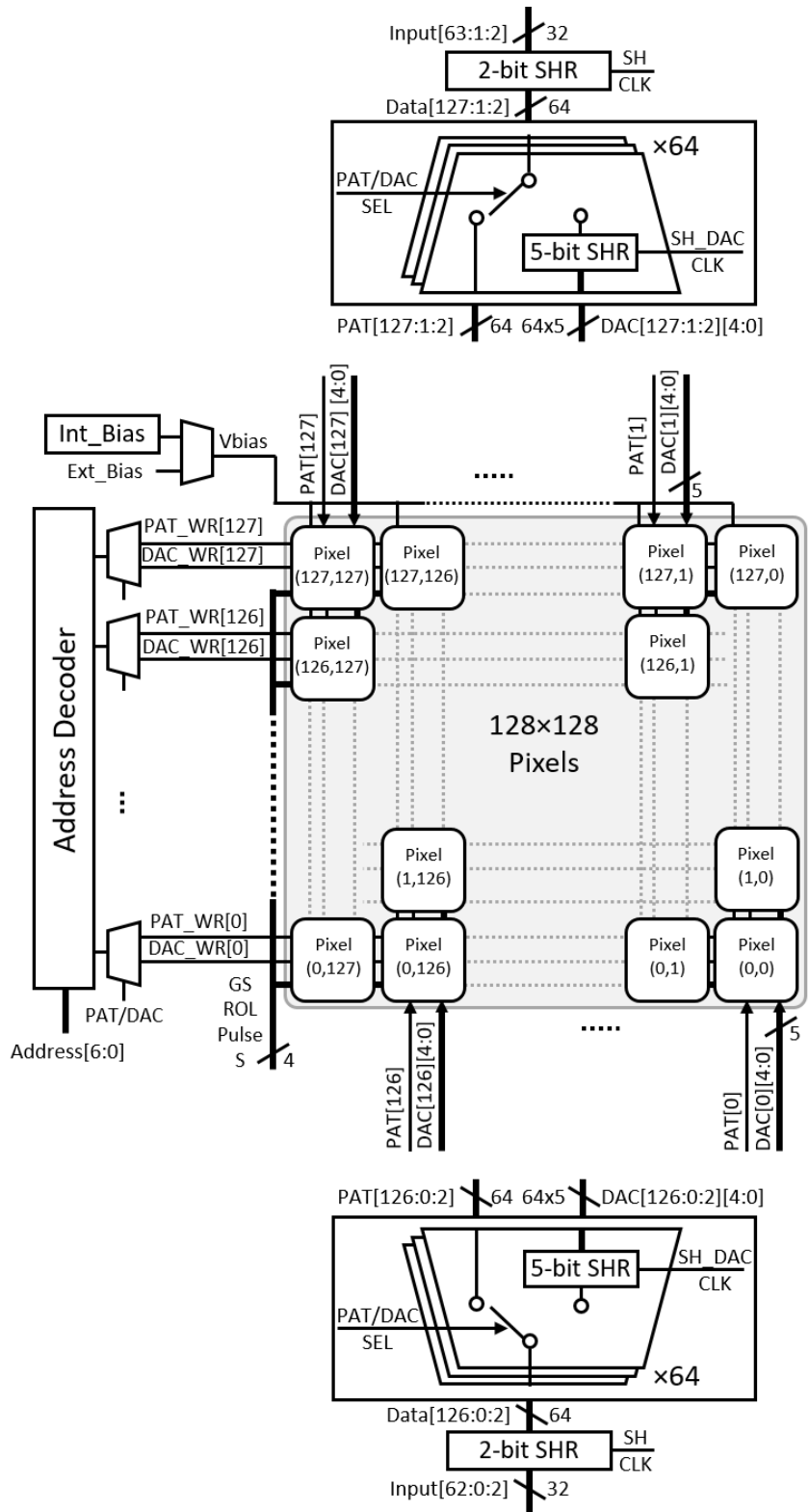

*Figure S1: Schematic diagram of the signal distribution circuit of the CMOS driver chip.*

The CMOS driver consists of three main parts: data communication, logic control and pixels.

*Data Communication*

There are 64 individual inputs, Input [63:0], accessible from the top and bottom of the chip. The input data is shifted for two clock pulses to generate Data [127:0] required to write in 128 pixels. According to PAT/DAC SEL, the data is used as pattern data, PAT [127:0], to write in memory cells or to generate 5-bit DAC values, DAC [127:0] [4:0].

*Logic Control*

The data writing into the pixels is performed on the rows sequentially, and on the columns simultaneously. The address decoder selects a row using Address [6:0] and then according to PAT_WR [127:0] and DAC_WR [127:0], the produced PAT [127:0] and DAC [127:0] [4:0] will be written in to 128 pixels (columns) in the selected row. The data write/display is controlled by GS and ROL inputs, globally. The inputs Signal and S are also applied globally to pulse the LEDs and drive them with higher level of currents. Figure S2 shows the timing diagram of the control signals in the array.

*Pixel*

Figure 3 of the main article shows the circuit diagram of a pixel which consists of a micro-LED bump-bonded to its CMOS driver through the anode contact. The pixel driver includes 2-bit memory, 5-bit current DAC, 2mA driver, PWM logic, and output stage for driving the micro-LED. The in-pixel memory stores the LED state to have global illumination of the whole array. The 2-bit memory enables writing a new data in one memory cell (e.g. M1) while the previous data from the other memory cell (e.g. M2) is displayed on the LED. This mode is achievable with ROL=0 while the data (PAT) can be written/displayed with both GS=0 and GS=1. In the rolling mode with ROL=1, the memory will be single bit and data will be rolled. The driver inverter drives the LED with the current generated by a 5-bit current DAC at 5V. The bias voltage of the DAC, Vbias, is generated by the internal bias, Int_Bias, to produce 0-75uA current range for the DAC. To produce different current range for the DAC, a different Vbias can be generated through the external bias, Ext_Bias, while the internal bias is disabled. In addition, a single transistor can drive the LED with higher level of current (2mA) while applying PWM with Pulse input. This allows using the array for applications which need higher level of optical power with narrow pulse width. Although a global reset is used for initializing the chip, with Pulse=0 can ensure the output driver is OFF during the chip power up.

Figure 4 in the main article shows a timing diagram of the data writing/displaying in a pixel. The DAC data is first generated when DAC_WR is active. Then the data pattern (PAT) is written to a memory cell while PAT_WR is active. When a new pattern is being written in M1 (or M2) the pattern data in M2 (or M1) is displayed on the LED.

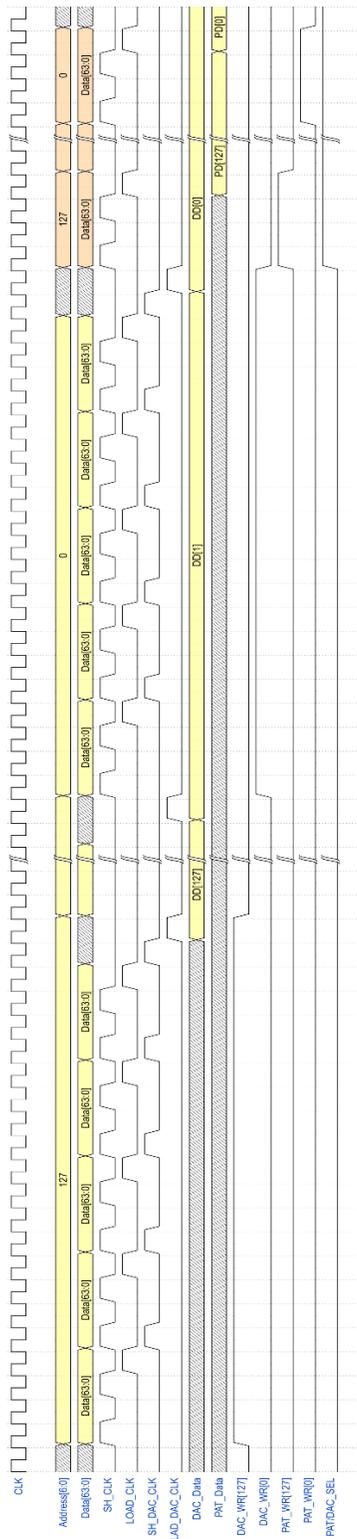

*Figure S2: Detailed timing diagram of gray scale and binary pattern loading.*

## 7. Commercial devices included in the literature survey

Figure 1 in the main article shows the framerate of digital light projectors versus the number of pixels. The data displayed in this figure were taken from the scientific publications referenced in the main article and from the data sheets of the following commercially available devices:

    Thorlabs EXULUS-4K1
    Thorlabs EXULUS-HD1
    Holoeye - PLUTO-2.1
    Hamamatsu X15213
    Santec - SLM-300
    Santec - SLM-200
    Holoeye GAEA-2
    Hamamatsu X10468
    Hamamatsu X13138
    Jasper JD2704
    Jasper JD955B
    Jasper JD7515
    Jasper JD7556VH
    TI DLP 9000
    IRIS AO PTT489
    IRIS AO PTT939
    BMC 4K-3.5-DM
    BMC Hex 1011
    BMC Hex 3063
    Samsung: the wall IW008J
    Samsung: the wall IW012J
    Samsung: the wall IW016J
    Samsung: the wall IW008R

    We also included information available from the Fraunhofer Institute for Photonic Microsystems IPMS on their spatial light modulator technology.

## 8. High rate pattern projection experiments

In order to demonstrate that a frame rate of 0.5 Mfps for binary patterns and 83 kfps for grayscale patterns is indeed achieved, the signal "LOAD_CLK" was monitored while loading a single pattern. The signal "LOAD_CLK" is a digital signal that alternates between logic high and low for the entire duration of the pattern load process and is continuously logic low while no pattern is loaded. Figure S3(a) shows the trace of LOAD_CLK while a single binary pattern is loaded, and it can be seen that the process is completed within 2 µs. Similarly, Figure S3(b) shows loading of a complete gray scale pattern within 12 µs. The total amount of data transferred for a single grayscale pattern is 5 times that of a single binary pattern. However, the loading time for a grayscale pattern is slightly longer than 10 µs due to the overhead required for the signals "SH_DAC_CLK" and "LOAD_DAC_CLK" which are illustrated in Figure S2.

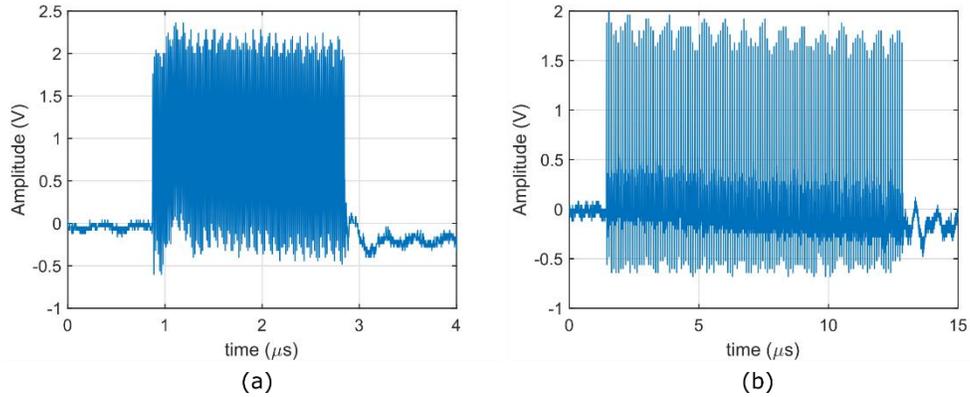

*Figure S3: Oscilloscope traces of the signal "LOAD_CLK" in Figure S2 when (a) loading a binary pattern and (b) loading a grayscale pattern.*

The camera communications experiment in section 3.1 of the main article used the Photron UX100 camera at a framerate of 0.8 Mfps. At this rate, the resolution of the camera is reduced to 8×640 pixels and thus only a subsection of the LED array can be monitored at a time. Three subsections were randomly selected for communications at 0.25 Mfps, and one of these subsections was used for communications at 0.4 Mfps. In all subsections we consistently see that the majority of bonded pixels have a bit error rate (BER) below the sensitivity of our measurement. This can be seen by the histograms shown in Figure S 4, which refer to one of the subsections. In the intensity histogram Figure S 4(a), we see that about 18% of the pixels have failed to bond correctly and do not light up, and a small number of pixels do light up albeit at a lower intensity than the majority of pixels. Figure S 4(b) plots the BER at 0.25 Mfps against the time-averaged pixel intensity. It can clearly be seen that that the vast majority of LEDs with a BER of less than the 7%-FEC threshold of $3.8 \times 10^{-3}$ have an intensity of less than 6000 camera ADC units. Only 5 pixels with BER below FEC threshold had a higher intensity, and it was found that these 5 pixels were all neighboring each other and had strong cross-talk, which we attribute to the pixels being electrically interconnected to each other during the bonding process. Figure S 4(c) shows the BER histogram at 0.25 Mfps. We see that the unbonded pixels have a BER of 0.5, and more than 76% of the pixels have a BER of $10^{-4}$ or less, i.e. below the sensitivity of the experiment, and more than 77% of the pixels have a BER below the 7%-FEC theshold. Comparison to Figure S 4(b) verifies that the pixels with BER between 0.5 and $10^{-4}$ are almost exclusively pixels with a low brightness in Figure S 4(a) except for the 5 interconnected pixels. Figure S 4(d) shows a similar BER histogram at 0.4 Mfps. Due to the marginal eye opening at this rate, the BER is more sensitive to the signal-to-noise ratio and thus the pixel intensity. Therefore, a lower percentage of 62% of the pixels achieved a BER below the sensitivity of the experiment, and more than 71% of the pixels achieved a BER below the 7%-FEC threshold.

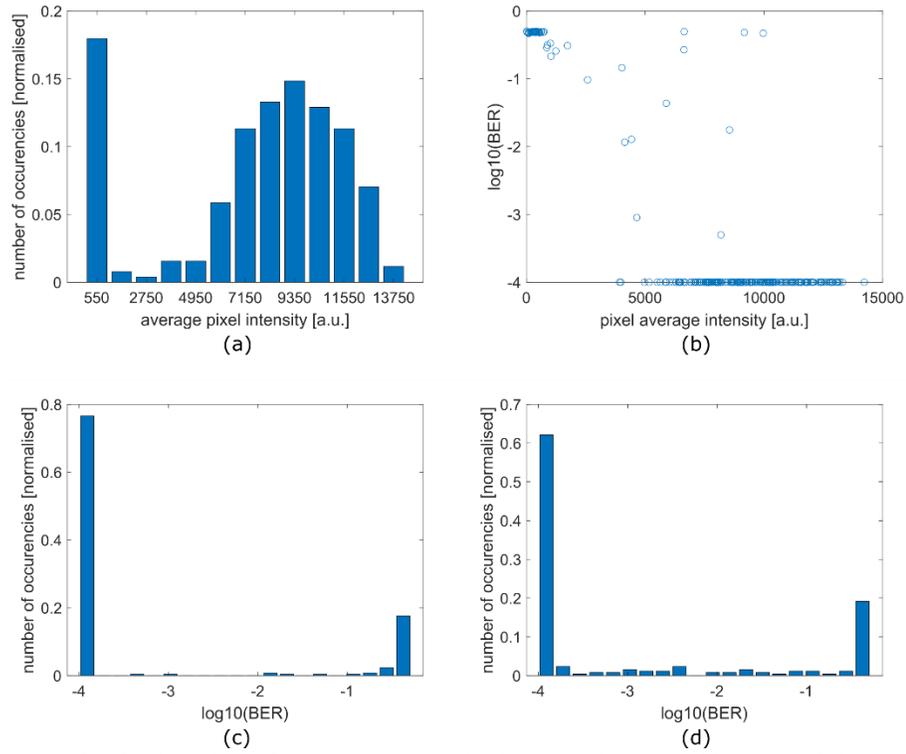

*Figure S 4: Statistics from the camera communications experiment: (a) histogram of time-average pixel intensity, (b) BER at 0.25 Mfps versus time-average pixel intensity, and histograms of BER at (c) 0.25 Mfps and (d) 0.4 Mfps.*